\documentclass[8.5pt,twoside,twocolumn]{article}
\usepackage[super,sort&compress,comma]{natbib} 
\usepackage[version=2]{mhchem}
\usepackage{times,mathptmx}
\usepackage{secsty}
\usepackage{balance} 
\usepackage{comment}
\usepackage{graphicx} 
\usepackage{lastpage}
\usepackage{float} 
\usepackage{wrapfig}
\usepackage{flushend}
\usepackage{subfigure} 
\usepackage[format=plain,justification=raggedright,singlelinecheck=false,font=small,labelfont=bf,labelsep=space]{caption} 
\usepackage{fancyhdr}
\begin{document}

\thispagestyle{plain}
\fancypagestyle{plain}{
\renewcommand{\headrulewidth}{1pt}}
\renewcommand{\thefootnote}{\fnsymbol{footnote}}
\renewcommand\footnoterule{\vspace*{1pt}%
\hrule width 3.4in height 0.4pt \vspace*{5pt}} 
\setcounter{secnumdepth}{5}

\makeatletter 
\def\subsubsection{\@startsection{subsubsection}{3}{10pt}{-1.25ex plus -1ex minus -.1ex}{0ex plus 0ex}{\normalsize\bf}} 
\def\paragraph{\@startsection{paragraph}{4}{10pt}{-1.25ex plus -1ex minus -.1ex}{0ex plus 0ex}{\normalsize\textit}} 
\renewcommand\@biblabel[1]{#1}            
\renewcommand\@makefntext[1]%
{\noindent\makebox[0pt][r]{\@thefnmark\,}#1}
\makeatother 
\renewcommand{\figurename}{\small{Fig.}~}
\sectionfont{\large}
\subsectionfont{\normalsize} 

\fancyfoot{}
\fancyfoot[RO]{\footnotesize{\sffamily{1--\pageref{LastPage} ~\textbar  \hspace{2pt}\thepage}}}
\fancyfoot[LE]{\footnotesize{\sffamily{\thepage~\textbar\hspace{3.45cm} 1--\pageref{LastPage}}}}
\fancyhead{}
\renewcommand{\headrulewidth}{1pt} 
\renewcommand{\footrulewidth}{1pt}
\setlength{\arrayrulewidth}{1pt}
\setlength{\columnsep}{6.5mm}
\setlength\bibsep{1pt}

\twocolumn[
  \begin{@twocolumnfalse}
\noindent\LARGE{\textbf{The Kinetic Energy of Hydrocarbons as a Function of Electron Density and  Convolutional Neural Networks$^\dag$}}
\vspace{0.6cm}

\noindent\large{\textbf{Kun Yao,$^{\ast}$\textit{$^{a}$} and John Parkhill,\textit{$^{a\ddag}$}}}\vspace{0.5cm}

\vspace{0.6cm}

\noindent \normalsize{We demonstrate a convolutional neural network trained to reproduce the Kohn-Sham kinetic energy of hydrocarbons from electron density. The output of the network is used as a non-local correction to the conventional local and semi-local kinetic functionals. We show that this approximation qualitatively reproduces Kohn-Sham potential energy surfaces when used with conventional exchange correlation functionals. Numerical noise inherited from the non-linearity of the neural network is identified as the major challenge for the model. Finally we examine the features in the density learned by the neural network to anticipate the prospects of generalizing these models.}
\vspace{0.5cm}
 \end{@twocolumnfalse}
  ]


\footnotetext{\textit{$^{a}$  Department of Chemistry and Biochemistry, University of Notre Dame, Notre Dame, IN 46556, jparkhil@nd.edu}}

\section{Introduction}
The ground state energy is determined by electron density ${n(r)}$\cite{PhysRev.136.B864}. However the overwhelming majority of 'DFT' applications use the Kohn-Sham(KS) formalism which instead yields the energy as a functional of a non-interacting wave-function\cite{PhysRev.140.A1133}. KS-DFT incurs a significant computational overhead in large systems because the minimal Kohn-Sham electronic state uses at least one wavefunction for each electron. In orbital-free (OF) DFT only one function, the density $n(r)$, is needed. Indeed, existing OF software packages are able to treat systems roughly an order-of-magnitude larger than efficient Kohn-Sham implementations on modest computer hardware\cite{Hung:2010qy,Xiang:2014aa,PhysRevB.92.014104}. The total energy in OF-DFT can be written as,
\begin{align}
   E^{OF-DFT}_{total}=T[n(r)]+E_{nu-el}[n(r)]+\\ \notag E_{hartree}[n(r])+E_{xc}[n(r)]+E_{nu-nu}
\end{align}
Because inexpensive density functionals are known for the other components of the electronic energy, one must only provide an accurate kinetic energy functional ($T[n(r)]$) to enjoy the computational advantages of OF-DFT\cite{Chai:2009aa}. In this paper we follow a totally naive and empirical route\cite{Wellendorff:2014aa} to approximations of $T[n(r)]$, based on convolutional neural networks we call CNN-OF\cite{Fukushima:1980aa,lecun1998gradient,Simard:2003aa,Matsugu:2003aa,Hinton:2006aa,Ciresan:2012aa}. Our approximation has useful accuracy, and it is able to predict bonding and shell-structure. It is designed to be compatible with existing Kohn-Sham exchange-correlation functionals and algorithms used to evaluate them. The functional is non-local and does not invoke a pseudo-potential\cite{Huang:2008aa,Lehtomaki:2014aa}. We also examine the features of the functional to infer features of the kinetic energy. \\
\indent     Several groups have developed quantitative approximations to the Born-Oppenheimer potential energy surface (BO-PES) using Neural Networks and other machine learning techniques\cite{behler2007generalized,handley2010potential,khaliullin2011nucleation,lopez2014modeling,Rupp:2015aa,behler2011neural}. In terms of theoretical detail our functional lies between between Kohn-Sham DFT and these Machine-Learning approximations to BO-PES. CNN-OF has the advantage that it models a property of electron density rather than molecular geometry, and so it generalizes between molecules. It could also be used to predict density-dependent properties and produce density embeddings for Kohn-Sham. OF-DFT is positioned to become an inexpensive approximation to Kohn-Sham theory, and is promising in multi-resolution schemes\cite{Wesolowski:1993aa,PhysRevA.77.012504,Shin:2012aa,Laricchia:2014aa}.
\\
\indent Accelerating progress is being made towards accurate kinetic energy functionals; a complete review is beyond the scope of this paper\cite{yang1986various,lembarki1994obtaining,Karasiev:aa,Karasiev:2012aa,PhysRevB.75.205122,Chai:2004aa,iyengar2001challenge,chakraborty2011failure,perdew1992generalized,PhysRevB.81.045206}. Most kinetic functionals descend from the local Thomas-Fermi(TF) or semi-local von-Weizsäcker(VW) functionals, which are exact for uniform electron gas, and one orbital systems, respectively\cite{Chan:2001rw}. They can be written as,
\begin{align}
  T_{TF}=\int C_{TF}n(r)^{5/3}dr \qquad T_{VW}=\frac {1}{8}\int \frac {{|\nabla n(r)|}^2} {n(r)} dr
\end{align}
where $C_{TF}$ equals  $\frac{3}{10}{(3{\pi}^2)}^{2/3}$.
The modifications to these functionals can be roughly classified by their locality, the first group being semi-local approximations based on gradient information. The accuracy of modern generalized gradient approximation (GGA) kinetic functionals has remarkably reached $\sim1\%$ for atoms\cite{PhysRevLett.106.186406}. However, existing GGAs have many qualitative failures\cite{PhysRevB.91.045124}: they do not predict shell structure of the density, and often catastrophically fail to predict even the strongest chemical bonds. Given the large magnitude of the kinetic energy relative to the XC energy, even this performance is impressive.\\
\indent The second class are non-local functionals of the density, and we can distinguish two important sub-types: two-point functionals based on a relation between density response and kinetic energy\cite{PhysRevB.53.9509,PhysRevB.57.4857,PhysRevB.45.13196,PhysRevB.58.13465,PhysRevB.60.16350} and empirical functionals based on the kernel method (KM) of machine learning\cite{Snyder:2015aa,Vu:2015aa}. When combined with pseudo-potentials, the non-local functionals usefully predict bulk properties of metals and semiconductors\cite{PhysRevB.86.235109}. Modifications such as angular momentum dependence further improve the accuracy of these functionals\cite{PhysRevLett.111.066402}. However to our knowledge there is limited evidence that the two-point functionals are practical for strongly inhomogeneous organic or biological material, and they depend on a pseudo-potential to avoid core electrons\cite{Xia:2012aa,PhysRevB.91.035126}. Another class of non-local functional recently appeared, empirical kinetic functionals based on the Kernel Method (KM) of Machine Learning\cite{snyder2012finding}. Ground-breaking studies of kernel method functionals have been restricted to 1-dimensional models of molecules, but have demonstrated several promising features, including bonding behavior\cite{Snyder:2013aa}. \\
\indent     This work is related to the KM approach, but makes several different design choices: 
\begin{itemize}
\item Our functional takes the form of an enhancement function, $F\{r,n(r')\}$, for a hybrid of the TF and VW functionals, and is locally integrated like an ordinary GGA xc-functional, although $F$ is non-local.
\item We use Convolutional Neural Networks (CNN's) rather than KM to learn the enhancement functional. 
\item Like the KM functionals, but unlike the two-point functionals CNN-OF is evaluated in real-space, and no pseudopotential (PS) is required. 
\item Our functional targets the positive semi-definite, non-interacting Kohn-Sham kinetic energy density $\tau_+$\cite{Sim:2003aa}. 
\end{itemize}
We explain the motivation and impact of each of these design choices in the remainder of this paper. The fourth choice constitutes an approximation, also made by Kohn-Sham calculations, that the kinetic contribution to correlation is negligible\cite{Whittingham:2005aa}. For our OF functional to be practically useful, it must be compatible with existing KS functionals, and so this approximation is a useful expedient that could be relaxed in future work. Note that there is no unique kinetic energy density\cite{Sim:2003aa,anderson2010ambiguous}; modeling $\tau_+$ has practical advantages discussed below. \\
\indent     There are known conditions of the exact XC-functional that are not satisfied by the approximate XC functionals\cite{perdew2014gedanken} in common use today. Likewise there are known features of the exact kinetic functional that are not satisfied by this work\cite{levy1988exact} including density scaling relations\cite{chan1999kinetic,Borgoo:2013qp}, response relations and asymptotic limits\cite{Lee:2009aa}. We expect that enforcing these conditions will be key to the future development. Within our scheme it is a simple matter to enforce these physical constraints with data that reflects the constraint, just as rotational invariance is enforced in image classification by generating rotated versions of input data\cite{Hinton:2006aa}. We defer a more complete investigation of these constraints to other work, and acknowledge from the outset that our functional is neither exact, nor unique defined, even based on its training set. This paper simply establishes that convolutional neural networks are able to predict the Kohn-Sham kinetic energy in real molecules from the density. 
\begin{figure}
\includegraphics[width=0.5\textwidth]{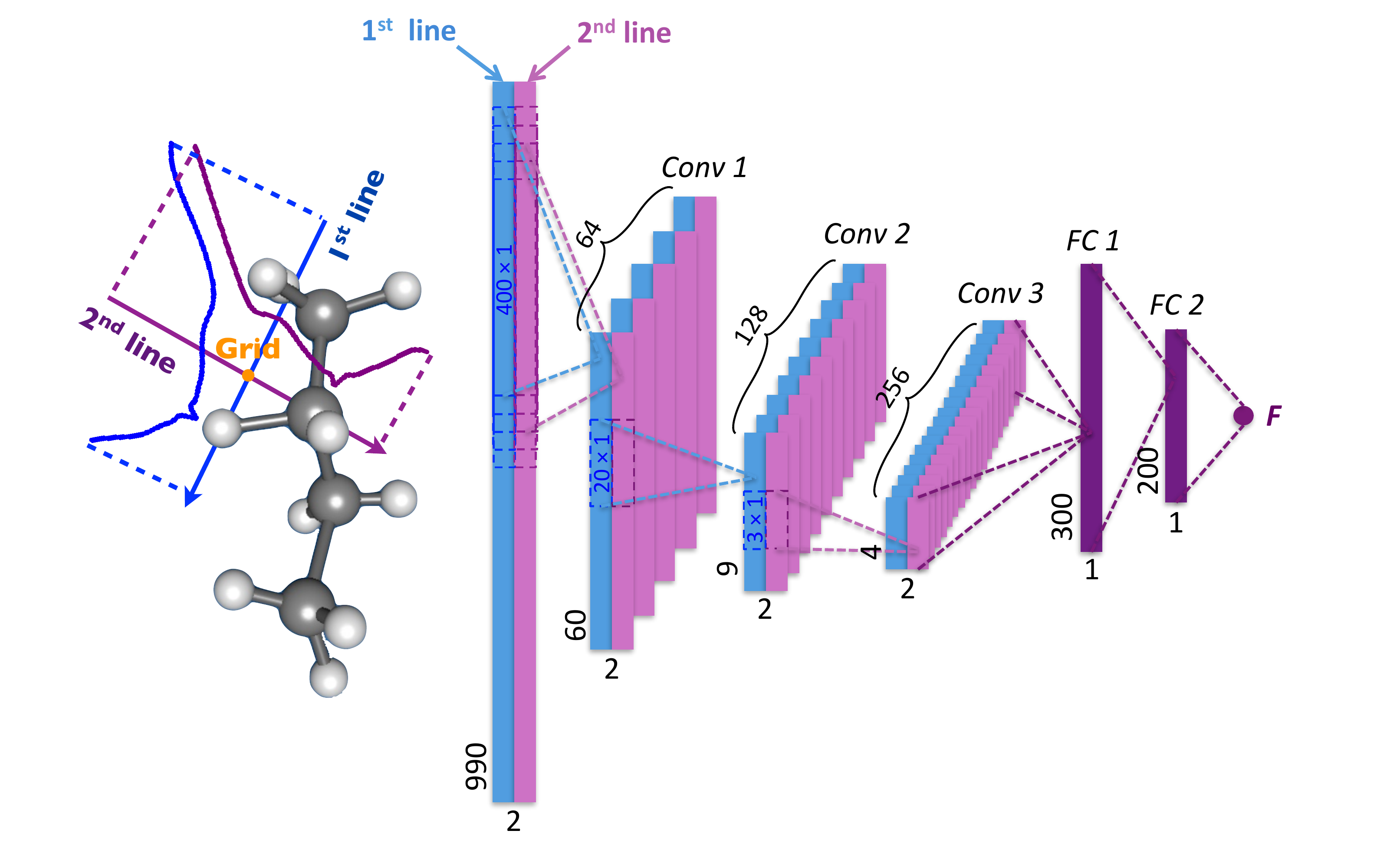}
\caption{The orange dot is the quadrature point ($r'$) at which the functional is being evaluated. The two lines which sample the normalized reduced gradient intersecting this point are input to the network which consists of three convolutional and two fully connected layers. The final layer outputs $F(r')$.}
\label{fig:Convlayers}
\end{figure}
\section{Functional Form}
\indent Like most exchange correlation functionals the form of our non-local kinetic energy is based on a local kinetic energy density that is exact for a physical model system (VW or TF) multiplied by a unitless non-local enhancement functional $F$, defined as:
\begin{align}
F_{tf(vw)}\{r,n(r')\} = {\tau_{+}(r)} / {\tau_{tf (vw)}(r)}
\end{align} 
where $n(r')$ is a sample of density near position $r$ centered there, $\tau_{+}(r) = \frac{1}{2}\sum_i |\nabla \phi_i(r)|^2$ is the positive-semidefinite Kohn-Sham kinetic energy density and $\tau_{tf(vw)}$ is the local TF or VW kinetic energy density\cite{anderson2010ambiguous}. The total kinetic energy can be written as:
\begin{align}
  T\{n(r)\}=\int F_{tf(vw)}\{r,n(r')\}\tau_{tf(vw)}(r) dr
\end{align}
In principle, the total kinetic energy should be the exact interacting kinetic energy of the whole molecule, but in practice the non-interacting kinetic energy is used in Kohn-Sham calculations. Our goal is to produce a kinetic functional which is compatible with existing KS functionals. We chose $\tau_{+}$ since it is everywhere positive, which avoids numerical problems, and the ratios of $\tau_{tf}$ and $\tau_{vw}$  to $t_+$ are well-behaved functions. Ideally, $F_{tf}$  should equal to 1 for any constant function input and $F_{vw}$ also equal to 1 for any one-orbital system\cite{garcia2012generalized}. We have examined several different choices of $F_{tf}$ vs $F_{vw}$ as described later on.

\subsection{Convolutional Neural Networks}
\indent  If the set of orbitals generating the density are known, calculating the non-interacting kinetic energy is trivial. This observation suggests that $T[n(r)]$ can be thought of as recognizing orbitals leading to the density, and that the most robust available statistical models for recognition are a logical choice for kinetic energy functionals. CNNs have emerged over the past few decades as the most powerful models for classification of image data. Previous 1D machine-learning work has employed a KM, kernel ridge regression (KRR), to learn the kinetic energy functional\cite{snyder2012finding,Snyder:2013aa,li2014understanding,Vu:2015aa}. The main strength of the KM relative to a CNN is a straightforwards deterministic learning process. The main drawback is difficulty scaling to large amounts of data with high dimensionality\cite{kmvsNN}. Neural networks (NNs), and in particular convolutional neural networks, are known for their ability to digest high dimensional data and vast data sets. The universal approximation theorem\cite{hornik1991approximation} shows that neural networks are capable of approximating \emph{arbitrary} functions on compact subspaces, a flexibility gained for the price of non-linearity\cite{larochelle2009exploring}.\\
\indent NNs are compositions of vector-valued functions separated into layers of neurons. Each layer linearly transforms a vector of input ($y$) with vectors learned parameters (weights $w$, and biases $b$). A non-linear activation function ($f$) is then applied to the result and yields the output value for the neuron, for example:\begin{align}
    y_{m}^{i}=f(b^{i}+\sum_{n}y_{n}^{i-1}w_{nm}^{i-1,i}) 
    \label{eq:nn}
\end{align}
where $y_{m}^{i}$ is the value of the neuron $m$ in layer $i$, $f$ is the non-linear activation function, $b^{i}$ is the bias of layer $i$, $w_{nm}^{i-1,i}$ is the weight of the connection. The activation function used in this work is the rectified linear unit (ReLU):$f(x) = \max(0, x)$. Our neural network consists of more than five layers which bear parameters (Fig. \ref{fig:Convlayers}). The weights and biases of the network are 'learned', by minimizing the prediction error of the network over a training set by gradient descent\cite{lecun1998gradient}.\\
\begin{figure}
\centering
\includegraphics[width=0.4\textwidth]{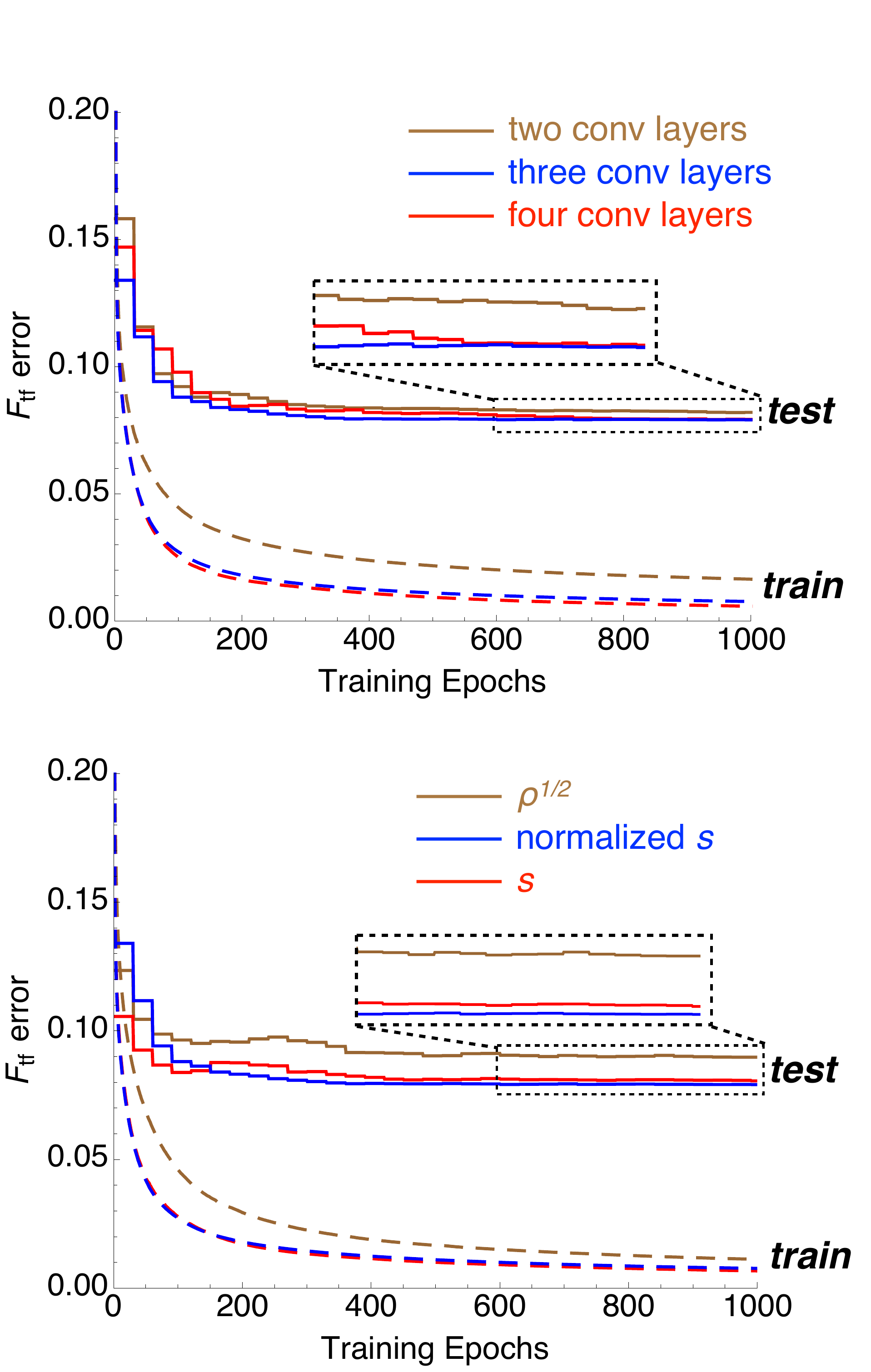}
\caption{Top: $F$ prediction errors (unitless) of CNNs with different number of convolutional layers for valance region as a function of learning epochs. An epoch is the number of gradient steps taken for the whole training set. CNNs with more convolutional layers produce less training error. However, the test error is saturated with three layers. Bottom: Learning curves of CNNs with different input types. A CNN using $\rho^{1/2}$ as input has the less accuracy than a CNN using normalized $s$ as input.}
\label{fig:learning_curve_2.pdf}
\end{figure}
\balance
\indent Convolutional neural networks are a constrained form of NNs, inspired by the structure of the animal visual cortex\cite{hubel1968receptive}. They are appropriate for data like images (and electron density) which have hierarchical local structure. They improve NNs by eliminating redundant parameters in the model. Each convolutional layer, as shown in Fig. 1, contains certain number of 'filters' which are local collections of neurons with fewer weights than inputs. These filters are convolved with the input data to produce output, for a filter size $p \times q$ the result is:
\begin{align}
y_{mn}^{i} = f(b^{i}+\sum_{a=0}^{a=p}\sum_{b=0}^{b=q}y_{(m+a)(n+b)}^{i-1}w_{ab}^{i-1,i})
\end{align}
where $y_{mn}^{i}$ is the value of neuron in layer i at position (m,n), $w_{ab}$ is the weight matrix of the filter. Since the weight matrices $w_{ab}^{i-1,1}$ are shared across area in convolutional layers, the number of parameters in a convolutional model is significantly reduced relative to a simple network. Each filter learns a separate sort of feature, and several filters are used in a layer which adds a new index of summation into Eq. \ref{eq:nn}. Obtaining the network's output involves a series of tensor contractions reminiscent of the coupled-cluster equations\cite{Parkhill:2010mw}. Both models derive flexibility by allowing non-linear dependence on parameter vectors. Convolutional layers are able to distill structure and improve the robustness of the NNs for object recognition.\cite{krizhevsky2012imagenet,lecun2015deep}
\subsection{Choices of Input and Network Shape}
\indent The whole density is an impractical amount of information to feed to $F$. Compact samples of $n(r)$ are theoretically enough\cite{mezey1999holographic}, but functionals based on small samples must be numerically unstable. While designing our functional we looked at the shape of $F$ for several small molecules, and also experimented with several sorts of input. Based on the structure of the exact $F$, we allow $F$ to depend on two, one-dimensional lines of the density centered at $r'$. These lines are oriented towards the nearest nuclei to let the functional perceive nearby shell-structure. This choice of input is arbitrary, and should be refined in future work. The left panel of Fig. 2 shows the results of some experiments with different versions of the density. The dimensionless gradient along the two lines we mentioned above is the scheme used for our results, but we also experimented with the density, the square root of the density and other variations. The density itself does not display much shape and has a large dynamic range. The reduced gradient ($s(r) = \left | \nabla n(r) \right | / 2k_{F}(r)n(r)$)
is a better choice: is dimensionless and lies in a small range, which makes it suitable as CNN input. It clarifies shell structure\cite{zupan1997density} which makes it easier for the CNN to learn $F$.  Indeed, One can see from Fig. 2 that using $s$ as input results in a large improvement over using $\rho^{1/2}$ as input. Each vector of $s$ fed into the network is normalized using the following local response normalization function\cite{krizhevsky2012imagenet}:
\begin{align}
    n(s^x)=\frac{s^x}{(1+0.01\sum_{x'=x-5}^{x'=x+5}(s^{x'})^2)^{0.5}}
\end{align}
where $x$ is the position in the line. The normalized $s$ as input further improves the numerical stability of the network, and focuses the network on learning spatial features.\\
\indent The performance of CNNs depends on their layer structure but the layer design must be found by a combination of intuition and trial and error\cite{kavzoglu1999determining}. Small networks will not have enough complexity to learn their training data. Large networks which are under-determined by their training data will eventually begin to learn the distribution of training data instead of the desired features, spoiling their generality. The right panel in Fig. \ref{fig:learning_curve_2.pdf} shows the learning curves of three different CNNs with different number of convolutional layers. Predictably, a CNN with more convolutional layers has less training error, however the test error shows the actual performance of the CNN and one can see the test error of the CNN with three convolutional layers and the CNN with four convolutional layers is actually quite similar. Based on this test and the computational constraints of evaluating the network in our GPU-CNN code, we chose three convolutional layers as a production model. The performance of fully-connected neural network and KRR have also been tested and the result are shown in Table S-2. As one can see, CNN has smaller test error than both methods. The size of the CNN we settled on is summarized in Fig. 1 and Table S-1. \\
\begin{figure}[h]
\centering
\includegraphics[width=0.4\textwidth]{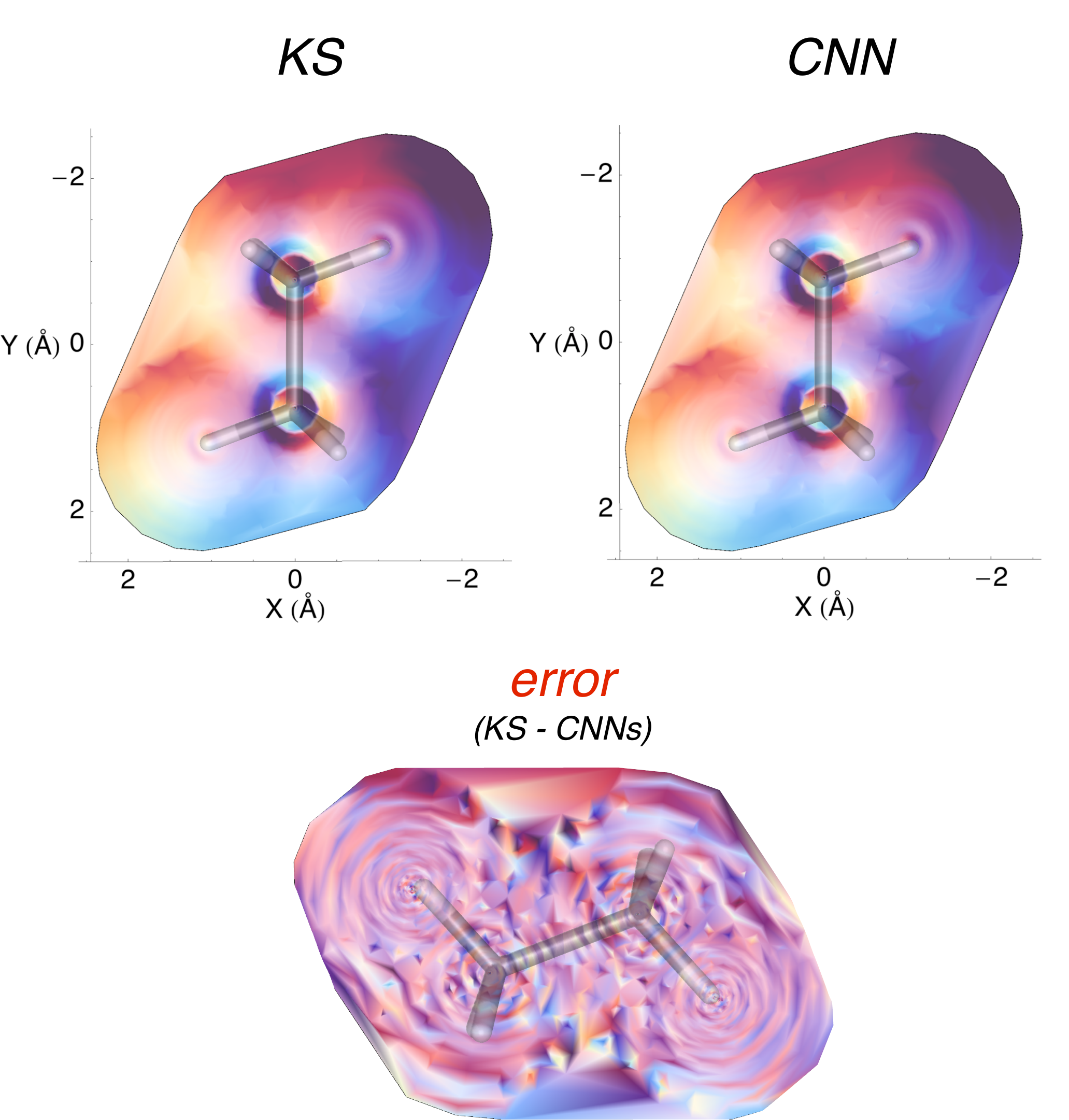}
\caption{Top left panel: Exact $F_{tf}$ calculated by Kohn-Sham method. Top right panel: $F_{tf}$ simulated by the trained neural network. Bottom panel: Error of the simulated $F_{tf}$ compared with the accurate $F_{tf}$. Neural networks simulate the exact $F_{tf}$ accurately, and structureless noise is the dominant error.}
\label{fig:Ftfplot}
\end{figure}
\indent In this paper, we focused on training kinetic energy functional for alkanes. The functional is trained based on its equilibrium structure and ten randomized structures of seven molecules: butane, 2,5-dimethylhexane, 2-ethyl-3,3-dimethylpentane, octane, tetramethylbutane, 2,3,3,4-tetramethylpentane and 5-ethyl-2-methylheptane. 2 \% of the standard xc-grid quadrature points of each structure were sampled as training samples, which are 3.7 Gigabytes of data. Each molecule is separated into three parts: a near carbon core region, valance region and tail region. Near carbon core region includes grids which lie less than 0.17 bohr from the nearest carbon nuclear and the learning target for grids in this region is $F_{tf}$. This region usually has less than 2 \% of the total grid points but contributes more than 20 \% of the total kinetic energy. Also considering the density in this region is chemically inert, it is treated separately . As mentioned above, $F_{tf}$ grows non-linearly at long range whereas $F_{vw}$ nicely converges to 1, so $F_{vw}$ was chosen as the learning target for grids in tail region which lie at least 2.3 bohr from the nearest nucleus. All the other grids consist of valance region, which is most sensitive to bonding and the learning target for grids in this region is $F_{tf}$. Each region was trained separately but with the same CNN architecture. To learn the functional we minimize sum-squared $F$-prediction errors over training densities as a function of the network's weights and biases. This error is itself a non-linear function, which we minimize from a random guess using stochastic gradient descent\cite{stograd} with analytic gradients provided by backpropagation\cite{Rumelhart:1986fk,krizhevsky2012imagenet}.\\
\indent The training was performed on Nvidia Tesla K80 GPU to accommodate the memory demands of the density input. $L2$ regularization was applied to prevent over fitting. Mini-batch stochastic gradient descent with momentum with batches of 128 samples was used during training. With this CNN structure and training scheme, the training wall time  for each part are shown in Table S-3 and the total training wall time is 29 hours. It is worth mentioning that under current learning scheme, our learning target is the kinetic energy enhancement factor instead of the kinetic energy, therefore the cancellation of the errors in the learning objective does not lead to the cancellation of the errors in the kinetic energy. This is the reason that the noisy error is uniformly distributed over grid points. These results could perhaps be further improved by manipulating the learning target.
\section{Results}
\indent     Our intended applications for these functionals are force-field like approximations to the BO-PES, and an inexpensive method to solve for densities of large systems. We note that because the wall-time cost of evaluating $F$ is constant regardless of grid size, and because the number of quadrature grid points in a molecule is linear with system size, this scheme is trivially linear scaling and naively parallel. We demonstrate that the functional learns properties of the kinetic energy, rather than merely the total energy by showing it's ability to predict $F$ throughout different regions of a molecule. We then show that the functional is accurate enough to predict bonding semi-quantitatively, and produces smooth PESs despite its non-linearity by examining its accuracy on describing the bonds in ethane and predicting kinetic energy along a KS molecular dynamics (MD) trajectory of 2-methylpentane. We conclude by examining the features learned by the CNN. Test molecules do not occur in the training set used to optimize the CNN, differing in both their bonding and geometry. A locally interfaced hybrid of modified BAGEL code\cite{Shiozaki:aa}, using the libXC library\cite{Marques:2012aa}, and Cuda-Convnet\cite{krizhevsky2012imagenet} was used to produce these results. A conventional pruned Lebedev atom-centered grid\cite{lebedev1975values} was used for integration of exchange-correlation energy and orbital-free kinetic energy in conjunction with Becke's atomic weight scheme\cite{becke1988multicenter}. The grid saturates the accuracy of most kinetic functionals to better than a microHartree. We note that in our hands $t_+$ saturates with grid more rapidly than typical XC functionals. The B3LYP exchange correlation functional\cite{becke1993density} and 6-31g* basis set was used in all results.  Q-Chem was also used to produce some comparison results\cite{shao2015advances}. Any quantity calculated with the CNN is obtained \emph{in single-precision arithmetic}; the fact that $\tau_+$ maintains sign is useful in this regard.\\
\subsection{Prediction of $F$}
\begin{figure}
\includegraphics[width=0.4\textwidth]{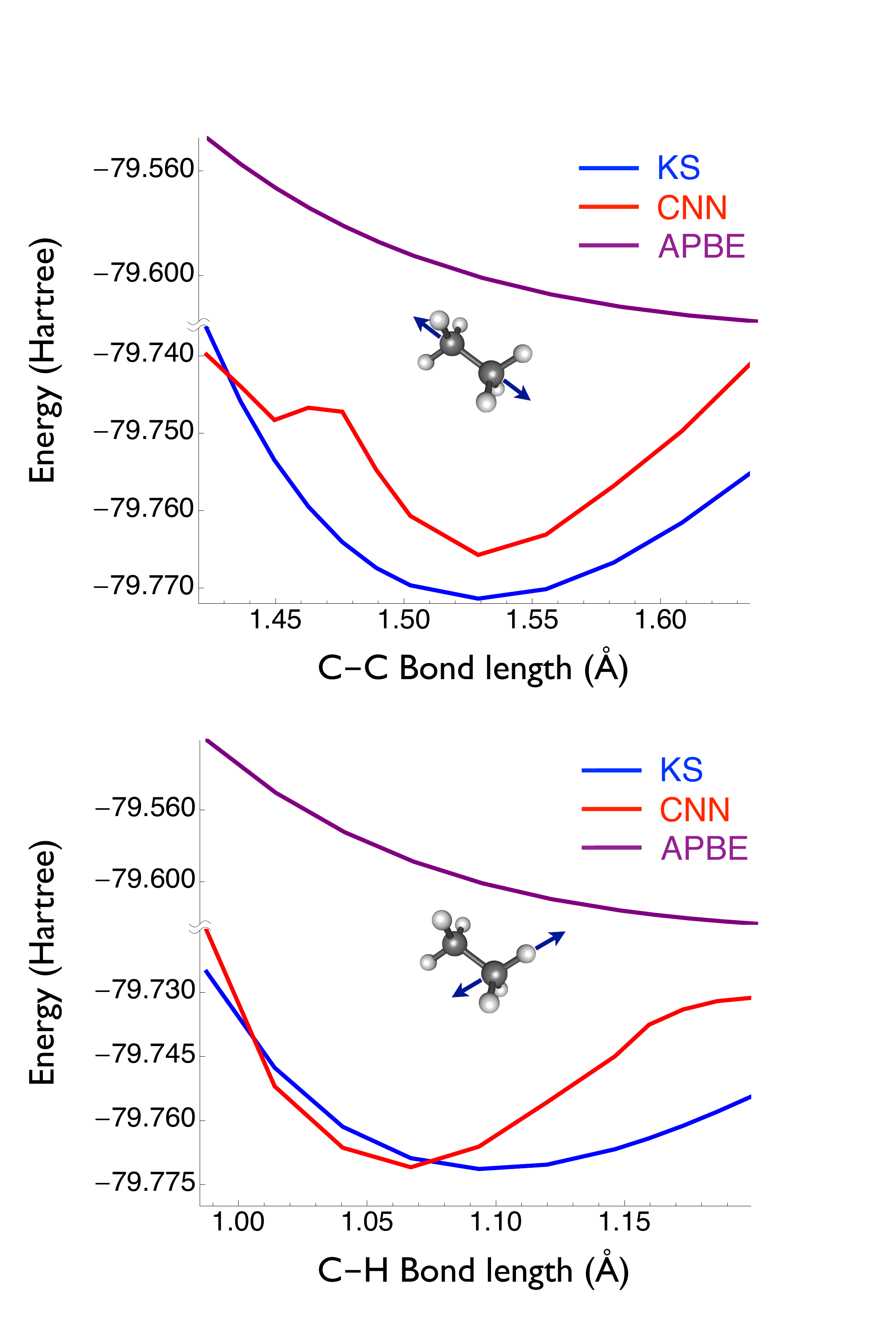}
\caption{Top panel: PES with kinetic energy calculated by Kohn-Sham, CNN and APBE kinetic energy functionals along the C-C bonding coordinate. Bottom panel: PES calculated with kinetic energy calculated by Kohn-Sham, CNN and APBE functionals along the C-H bonding coordinate. Our kinetic functional trained by CNN successfully find the local minimum with reasonable bond length}
\label{fig:CCCHbond.pdf}
\end{figure}
\indent The bonding curve test for ethane includes 9 images both for C-C bonding and C-H bonding, which contains 954,000 grid points in total.  The MD trajectory of 2-methypentane includes 17 steps which contains 3706,000 grid points in total. As mentioned above, There are 533,760 training samples each consisting of 2000 inputs, and there are more than 800,000 parameters in the CNN. The number of test samples that we predict outnumber the training samples by a factor of eight. 
\indent After training a CNN to reproduce the non-local enhancement factor, we wanted to establish whether there was any local trends in the accuracy of its prediction. To measure this we plot the $F$ produced by our learned model alongside the 'accurate' Kohn-Sham enhancement factor. The accurate $F_{tf}$ and the one generated by the trained CNN of ethane along the C-C-H plane is shown in Figure \ref{fig:Ftfplot}. As one can see, the simulated $F_{tf}$ is smooth and reproduces all the fine structure of the Kohn-Sham $F_{tf}$ surface, including the shell structure of carbon atom and the singularity of $F_{tf}$ near carbon core.  The root-mean-square deviation of $F_{tf}$ of ethane between accurate $F_{tf}$ and simulated one at its equilibrium structure is 0.03 (unitless). Considering the range of $F_{tf}$, between 0 and 13, the stochastic nature of our minimization, and the single precision arithmetic used, the CNN's $F_{tf}$ is remarkably accurate. The shape of the error is relatively uniform noise distributed throughout the volume of the molecule. This noisy error is the greatest challenge facing CNN-OF, especially because noise near the core where almost all the density lies can render the functional chemically inaccurate. This noise is intimately related to the non-linearity of the CNN. It is known from image recognition NN predictions are inherently unstable to infinitesimal perturbations \cite{2014arXiv1412.6572G}. One can imagine several remedies: solving for an ensemble of networks, training on adversarial examples, or denoising\cite{burger2012image}. We will show in the following sections that although noise is the dominant problem, it does not preclude chemical applications of CNN-OF. 
\subsection{Bonding and Potential Energy Surfaces}
\indent Poor prediction of chemical bonds and bond energies has kept OF-DFT out of the mainstream of computational chemistry, although the errors in existing kinetic functionals are relatively small. Generally the magnitude of the kinetic energy decreases as atoms are drawn apart and the failure to bond is simply due large contribution of kinetic energy, and a small error in the slope. As figure \ref{fig:CCCHbond.pdf} shows, the GGA-based APBE kinetic functional\cite{PhysRevLett.106.186406} (which is amongst the best kinetic functionals available in LibXC) fails to predict the C-H bond and C-C bond in ethane. The bonding curve generated with our trained kinetic energy functional successfully predicts local minimal both for C-C bond and C-H bond and the bonding curves are smooth especially in the vicinity of the minimum. Both the predicted C-C and C-H bond lengths lie within 50 milli-Angstroms of the KS value. 
\subsection{Accuracy along an MD trajectory}
\begin{figure}
\centering
\includegraphics[width=0.4\textwidth]{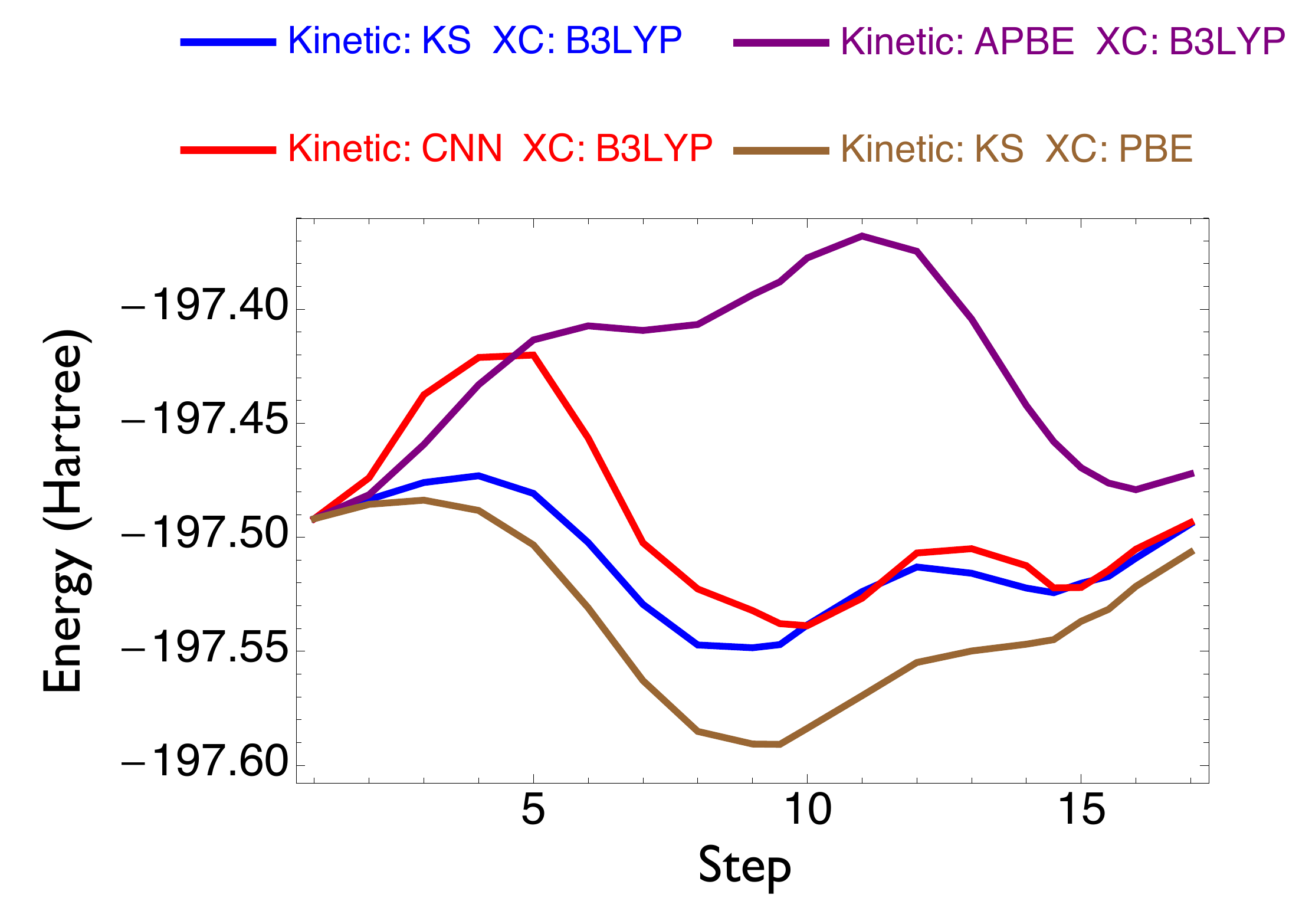}
\caption{PES with kinetic energy calculated by Kohn-Sham, CNN and APBE kinetic energy functionals along 17 steps of a Kohn-Sham molecular dynamics trajectory of 2-methypentane. For better comparison, the APBE and the CNN curves are translated so that they start at the same point as the curve. The PES calculated with CNN kinetic energy simulates the KS qualitatively, the relative errors are similar to the difference between hybrid B3LYP and GGA PBE.}
\label{fig:md_curve_with_APBE.pdf}
\end{figure}
\indent     A functional would be useless if it were only accurate in the vicinity of stationary geometries. To test generalization of our functional away from minimum, we examined a section of a high-temperature MD trajectory. Note, we have not implemented nuclear gradients of CNN-OF. This section uses a KS nuclear trajectory, and asks whether CNN-OF can produce a qualitatively correct surface for the KS geometries when several atoms are moving at once. Seventeen continuous steps of the molecular dynamics trajectory of 2-methypentane obtained at 1800 K are sampled. In our current implementation this is still a demanding task, because the densities of each point which are 10s of gigabytes of data are stored and processed. The CNN kinetic functional captures the general shape of PES including the positions of maximum and stationary points, although the curvature is imperfect. The error the CNN incurs is comparable to the error resulting from replacing B3LYP by PBE. Based on this trial and the previous bonding curves, we believe that CNN's are quite promising for the prediction of Kohn-Sham kinetic energies. 
\begin{figure*}
\centering
\includegraphics[width=0.8\textwidth]{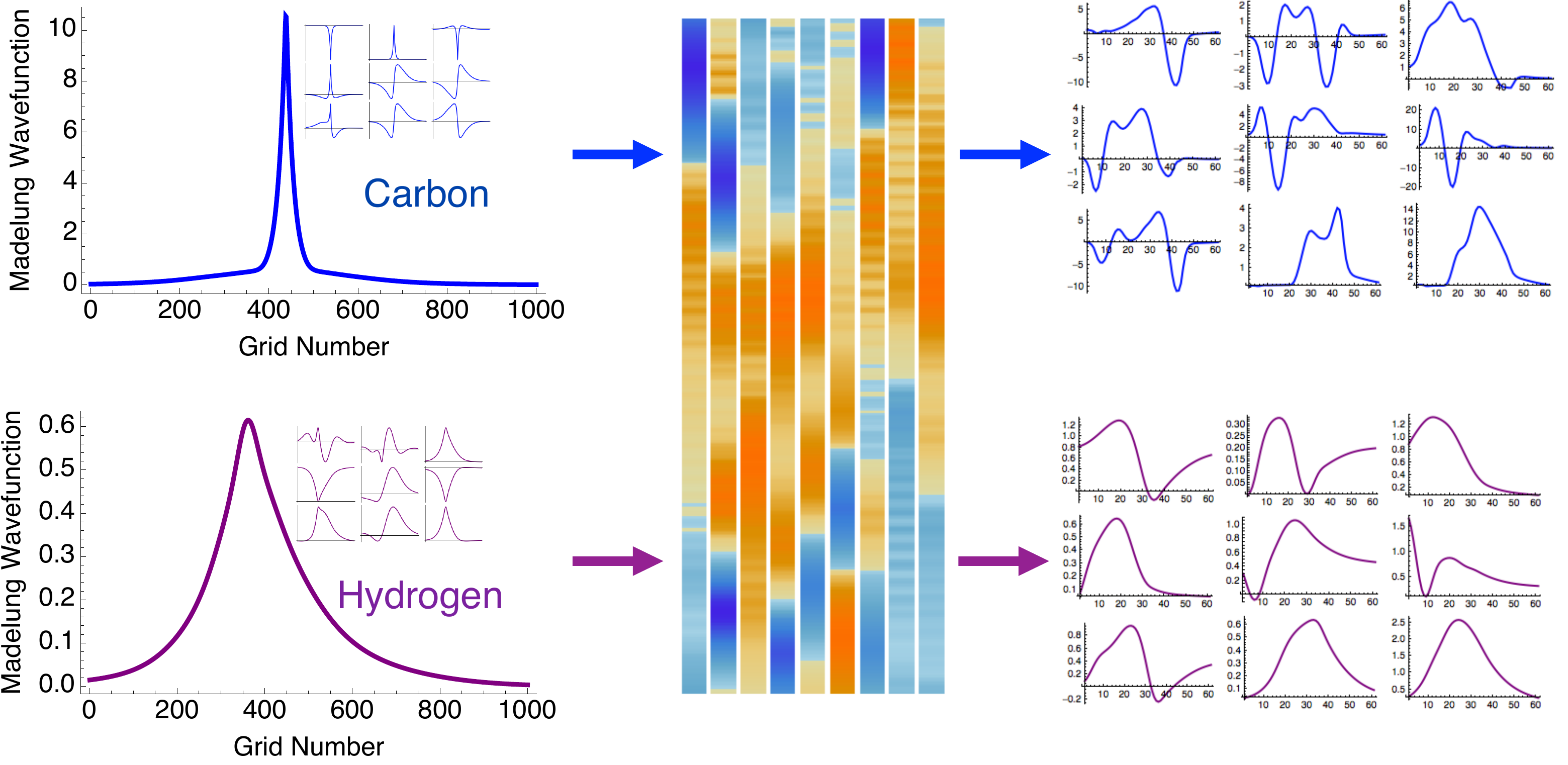}
\caption{Left panel: Madelung wavefunction (insert picture: molecular orbitals) along the sampled line of ethane (top: line points to carbon atom, bottom: line points to hydrogen atom). Middle panel: weights of the filters in first convolutional layer. Right panel: curves after the density is transformed by the filters. The transformed density near carbon atom shows obvious nodal structure, while the transformed hydrogen density does not.}
\label{fig:filtered_image_new.pdf}
\end{figure*}\balance
\subsection{What the Network Learns}
\indent     In order to gain some insights into the neural network we can examine the weights in the lowest convolutional layers. These weight vectors are 'features' recognized in the density by the CNN. For example in a network trained to identify images of people, the weights of low-lying convolutional layers look like patterns observed in images (edges and corrugation). The weight vectors of higher layers take on the shapes of increasingly complex objects in images (eyes, hands etc.). The smooth structure of the weight vectors is entirely due to the learning process because the network is initialized with random numbers. Features also tend to be nearly orthogonal, and by observing features we can diagnose over-parameterization, because excess filters do not lose their noisy character even when learning is complete. \\
\indent     For this experiment we trained a network on the Madelung wavefunction, $n(r)^{1/2}$, for simplicity. The lowest feature layers correspond directly to density of the molecule and The higher levels learn abstract representations of $\tau_+$ features that are difficult to interpret. Looking at the weight matrices of the lowest convolutional layer (Fig. \ref{fig:filtered_image_new.pdf}), the non-locality of the kinetic functional is superficially obvious. Most features extend several angstroms away from the point at which the enhancement is being evaluated. We can also infer that the real-space size of the sample we feed the network (10\AA) is adequate, since weights at the edges are near zero, and that the network has inferred some locality in $\tau_+$'s dependence on the density. The non-locality of the weights corresponds to the improvement of the two-point functionals over GGAs. We also see that we have basically saturated the number of features we can learn from our data. Small noisy oscillations persist in the weight vectors.  \\
\indent     Looking at the output of the lowest convolutional layer we can also see how the network is able to distinguish the shell structure of different atoms using its convolutional filters. We can see from figure \ref{fig:filtered_image_new.pdf} that even though the density along the line points to hydrogen atom and carbon atom has similar single peak shape, they become much more distinguishable after the transformation of the first convolutional layer. The when shown the sample from carbon's density the network produces outputs with many nodes, while the inputs pointing to hydrogen's density have none. Subsequent layers can easily detect these edges in their input to discriminate atoms. Ultimately to describe many materials, the network must learn the shell structures of several atoms, this basic classification is the first step in that process.  

\section{Discussion and Conclusions}
\indent We have shown that an empirically trained Convolutional Neural Network can usefully predict the Kohn-Sham kinetic energy of real hydrocarbons given their density. Our scheme can be practically integrated with existing functionals and codes without pseudo-potentials. The network is able to learn non-local structure and predict bond lengths despite being constructed in a naive way. We have shown (as predicted) that roughness brought about by the non-linearity of the network is the key challenge to further development, but that useful accuracy is already possible. There are several venues to improve on the model, for example by enforcing physical constraints, improving the numerical precision, and data used to train the network. For the foreseeable future OF-DFT will, at-best, be an inexpensive approximation to Kohn-Sham, and so the computational cost of evaluating the functional must be considered alongside these improvements. 

We thank The University of Notre Dame's College of Science and Department of Chemistry and Biochemistry for generous start-up funding, and Nvidia corporation for a grant of processors used for the work. 



\footnotesize{
\bibliography{OrbitalFree} 
\bibliographystyle{rsc} 
}

\end{document}